\documentclass[aps,prb,twocolumn,showpacs]{revtex4}

\usepackage{graphicx}
\def  \bsig    {\mbox{\boldmath$\sigma$}}

\def  \btau    {\mbox{\boldmath$\tau$}}

\begin{document}

\title{Macroscopic description of current induced switching due to spin-transfer}

\author{J. Barna\'s$^{1,2}$, A. Fert$^3$, M. Gmitra$^1$, I. Weymann$^1$, and V. K. Dugaev$^{4}$}
\affiliation{$^1$Department of Physics, Adam Mickiewicz University, Umultowska 85, 61-614~Pozna\'n, \\
$^2$ Institute of Molecular Physics, Polish Academy of Sciences,
M.~Smoluchowskiego~17, 60-179~Pozna\'n, Poland\\
$^3$Unit\'e Mixte de Physique CNRS/THALES associated with
Universit\'e Paris-Sud, Domaine de Corbeville, 91404 Orsay, France \\
$^4$Department of Physics and CFIF, Instituto Superior Tecnico,
Av. Rovisco Pais, 1049-001, Lisbon, Portugal; and Institute for
Problems of Materials Science, NASU, Vilde 5, 58001 Chernovtsy,
Ukraine}

\date{\today }

\begin{abstract}
\vskip0.2cm \noindent We develop a macroscopic description of the
current-induced torque due to spin transfer in a layered system
consisting of two ferromagnetic layers separated by a nonmagnetic
layer. The description is based on i) the classical spin diffusion
equations for the distribution functions used in the theory of
CPP-GMR, ii) the relevant boundary conditions for the longitudinal
and transverse components of the spin current in the situation of
quasi-interfacial absorption of the transverse components in a
magnetic layer. The torque is expressed as a function of the usual
parameters derived from CPP-GMR experiments and two additional
parameters involved in the transverse boundary conditions. Our
model is used to describe qualitatively normal and inverse
switching phenomena studied in recent experiments. We also present
a structure for which we predict only states of steady precession
above a certain critical current. We finally discuss the limits of
a small angle between magnetic moments of the ferromagnetic layers
and of vanishing imaginary part of the mixing conductance.

\end{abstract}

\pacs{75.60.Ch,75.70.Cn,75.70.Pa}

\maketitle

\section{Introduction}

The magnetic moment of a ferromagnetic body can be switched
without applying an external magnetic field, but only by transfer
of electron spins carried in a spin polarized current. The concept
of spin transfer has been introduced by Slonczewski
\cite{slonczewski96} and appears also in several papers of Berger
\cite{berger96}. Magnetic switching by a spin polarized current
has been now confirmed in extensive series of experiments
\cite{tsoi98,sun99,urazhdin04,darwish04}. Spin transfer is also an
important turning point in spintronics. In spintronic phenomena of
the first generation, like giant magnetoresistance (GMR)
\cite{baibich88} or tunneling magnetoresistance (TMR)
\cite{moodera95}, the magnetic configuration of a nanostructure is
detected by an electrical current. On the contrary, in spin
transfer experiments a magnetic configuration is created by a
current. This possibility of  back and forth magnetic switching
opens new fields for spintronics.

Current-induced magnetic switching (CIMS) has been clearly
demonstrated \cite{sun99} by experiments on structures F1/N/F2
consisting of two ferromagnetic layers of different thicknesses
separated by a nonmagnetic layer N. Starting from a parallel
configuration of the magnetizations in F1 and F2, a current
exceeding a certain critical value can reverse the magnetic moment
of the thinner magnetic layer to set up an antiparallel
configuration. In turn, a current in the opposite direction can
switch back the structure to the parallel configuration. With an
applied field, the spin transfer mechanism can also generate a
steady precession of the magnetization, detected by oscillations
of the current in the microwave frequency range \cite{kiselev03}.

In the concept introduced by Slonczewski \cite{slonczewski96}, as
well as in most theoretical models
\cite{weintal00,slonczewski02,stiles02}, the current-induced
torque acting on the magnetization of a magnetic layer is related
to the spin polarization of the current and, more precisely, to
the absorption of the transverse component of the spin current by
the magnetic layer. From CPP-GMR experiments one knows that the
spin polarization of the current is due to spin dependent
reflections at interfaces and to spin dependent scattering within
the magnetic layers. For CIMS, similarly, recent experiments
\cite{darwish04} have shown that the switching currents can be
modified and even reversed by doping the magnetic layers with
impurities of selected spin dependent scattering cross-sections.

Both CPP-GMR and CIMS depend on spin accumulation effects. This is
well known for CPP-GMR. For CIMS, this has been shown by
experiments in which the spin accumulation profile is manipulated
by introducing spin-flip scattering at different places in the
structure \cite{urazhdin04}. It turns out that both the GMR effect
and the spin transfer torque can be enhanced by introducing
spin-flip scattering outside a F1/N/F2 trilayer (in the leads) or
reduced by spin-flip scattering in the nonmagnetic layer N
\cite{urazhdin04}. This calls for a unified theory of CPP-GMR and
spin transfer torque, taking into account spin accumulation, spin
relaxation, and both interface and bulk spin dependent scattering.
This is actually the direction of most recent theoretical
developments \cite{slonczewski02,stiles02,brataas00}.

The model we present in this paper fits directly  with the
interpretation of CPP-GMR data in the model of Valet and Fert (VF)
\cite{valet93}. Most of the parameters of our description can be
derived directly from the analysis of CPP-GMR experimental data
\cite{bass99}, that is interface and bulk spin asymmetry
coefficients, interface resistance, layer resistivities, and spin
diffusion lengths. As we will see below, two additional
parameters, namely the real and imaginary parts of the mixing
conductance \cite{brataas00,brataas01}, are also needed. They can
be derived from quantum-mechanical calculations \cite{xia01} of
the transmission of spin currents at the interface under
consideration. By introducing into our model calculated values of
the mixing conductance in addition to the set of parameters
derived from GMR experiments, we calculate the current-induced
torque for different types of structures in order to understand
how its sign, amplitude and angular variation depend on the spin
asymmetry coefficients and spin accumulation effects. The
calculations of our model are based on macroscopic transport
equations similar to those derived from the Boltzmann equation
approach of the VF model \cite{valet93} for the CPP-GMR of
multilayers with collinear magnetizations. Stationary charge and
spin current are described by classical diffusion equations. We
assume that the absorption of the transverse component of the spin
currents is quasi-interfacial, as this has been justified by a
quantum description of the transmission of transverse spin current
into a ferromagnetic layer. This assumption allows to derive some
effective boundary conditions for the spin accumulation and spin
current drops at the interfaces \cite{brataas01}, and this way
also to calculate the torque acting on a magnetic film for an
arbitrary angle between magnetic moments of the two ferromagnetic
films in a structure. Thus, in a certain sense this extends to an
arbitrary angle a small-angle description{\cite{fert04} that has
been used for the interpretation of recent
results\cite{darwish04}.

The paper is organized as follows. Macroscopic equations
describing currents and spin accumulation inside the films are
derived in section 2. The boundary conditions and general formulae
for the torque in a four-layer structure are presented in sections
3 and 4, respectively. Numerical results for the structure with
two magnetic films are presented and discussed in section 5. The
limiting case of real mixing conductance is considered in section
6. The limit of a small angle between magnetizations is discussed
in section 7, whereas final conclusions are in section 8.

\section{Currents and spin accumulation inside magnetic and nonmagnetic films}

We assume the electrical current in the multilayer is carried by
free-like electrons of equal concentrations in all the layers and
without any spin polarization at equilibrium. The distribution
function $\check{f}$ inside the films is a $2\times 2$ matrix in
the spin space, and its spatial variation can be described by the
diffusion equation. We assume the distribution functions are
uniform in the plane of the films, and vary only along the axis
$x$ normal to the films. Let us consider first ferromagnetic
layers.

\subsection{Magnetic films}

The diffusion equation for arbitrary spin quantization axis takes
then the form \cite{brataas01}
\begin{equation}
\check{D}\frac{\partial^2 \check{f}}{\partial x^2} =
\frac{1}{\tau_{sf}} \left[ \check{f}-\check{1}\frac{{\rm
Tr}\{\check{f}\}}{2}\right],
\end{equation}
where $\check{D}$ is the diffusion $2\times 2$ matrix in the spin
space, $\check{1}$ is the $2\times 2$ unit matrix, and $\tau_{sf}$
is the spin-flip relaxation time. As it has been already mentioned
in the introduction, we assume that the internal exchange field
inside ferromagnetic metals is strong enough so that the component
of the distribution function perpendicular to the local
magnetization vanishes. Thus, the distribution function is
diagonal when the spin quantization axis is parallel to the local
spin polarization of the ferromagnetic system. Equation (1) can be
then written as
\begin{equation}
D_\uparrow \frac{\partial^2 f_\uparrow }{\partial x^2} =
\frac{1}{2\tau_{sf}} (f_\uparrow -f_\downarrow ),
\end{equation}
\begin{equation}
D_\downarrow \frac{\partial^2 f_\downarrow }{\partial x^2} =
\frac{1}{2\tau_{sf}} (f_\downarrow -f_\uparrow ),
\end{equation}
where  $f_\uparrow$ and $f_\downarrow$ are the distribution
functions for spin-majority and spin-minority electrons,
respetively.

The above system of two equations can be rewritten as
\begin{equation}
\frac{\partial^2 (f_\uparrow -f_\downarrow ) }{\partial x^2} =
\frac{1}{l_{sf}^2} (f_\uparrow -f_\downarrow ),
\end{equation}
\begin{equation}
\frac{\partial^2 (f_\uparrow +f_\downarrow ) }{\partial x^2} =
\eta \frac{\partial^2 (f_\uparrow -f_\downarrow ) }{\partial x^2},
\end{equation}
where
\begin{equation}
\frac{1}{l_{sf}^2}=\frac{1}{2}\left(\frac{1}{l_{\uparrow}^2}+\frac{1}{l_{\downarrow}^2}\right)
\end{equation}
with $l_\uparrow^2 =D_\uparrow \tau_{sf}$ and $l_\downarrow^2
=D_\downarrow \tau_{sf}$, and $\eta$ defined as
\begin{equation}
\eta = - \frac{D_\uparrow -D_\downarrow}{D_\uparrow
+D_\downarrow}.
\end{equation}

Equations (4) and (5) can be rewritten in terms of the
electro-chemical potentials ${\bar\mu}_{\uparrow }$
(${\bar\mu}_{\downarrow}$) for spin-majority (spin-minority)
electrons as
\begin{equation}
\frac{\partial^2 ({\bar\mu}_\uparrow -{\bar\mu}_\downarrow )
}{\partial x^2} = \frac{1}{l_{sf}^2} ({\bar\mu}_\uparrow
-{\bar\mu}_\downarrow ),
\end{equation}
\begin{equation}
\frac{\partial^2 ({\bar\mu}_\uparrow +{\bar\mu}_\downarrow )
}{\partial x^2} = \eta \frac{\partial^2 ({\bar\mu}_\uparrow
-{\bar\mu}_\downarrow ) }{\partial x^2}.
\end{equation}
The above equations are equivalent to the equations derived from
the Boltzmann equation approach by Valet and Fert \cite{valet93}.

Solution of Eqs (8) and (9) gives
\begin{displaymath}
{\bar\mu}_\uparrow  = (1+\eta )[A\exp(x/l_{sf})
\end{displaymath}
\begin{equation}
+ B\exp(-x/l_{sf})] +Cx+G ,
\end{equation}
and
\begin{displaymath}
{\bar\mu}_\downarrow  =  (\eta -1)[A\exp(x/l_{sf})
\end{displaymath}
\begin{equation}
+ B\exp(-x/l_{sf})] +Cx+G,
\end{equation}
where $A$, $B$, $C$ and $G$ are constants to be determined later
from the appropriate boundary conditions.

The electro-chemical potentials can be written as
\begin{equation}
\check{{\bar\mu}} = {\bar\mu}_0\check{1}  + g\check{\sigma}_z
\end{equation}
with
\begin{equation}
{\bar\mu}_0 = ({\bar\mu}_\uparrow +{\bar\mu}_\downarrow )/2
\end{equation}
and
\begin{equation}
g = ({\bar\mu}_\uparrow -{\bar\mu}_\downarrow )/2.
\end{equation}
Thus, the explicit forms for ${\bar\mu}_0$ and $g$ are
\begin{equation}
{\bar\mu}_0  =  \eta[A\exp(x/l_{sf})+ B\exp(-x/l_{sf})] +Cx+G,
\end{equation}
and
\begin{equation}
g  = A\exp(x/l_{sf})+ B\exp(-x/l_{sf}).
\end{equation}

For an arbitrary quantization axis the particle and spin currents
are given by the $2\times 2$ matrix $\check{j}$ in the spin space
\begin{equation}
\check{j}=-\check{D}\frac{\partial \check{f}}{\partial x}=-\rho
(E_F)\check{D}\frac{\partial \check{{\bar\mu}}}{\partial x},
\end{equation}
where $\rho (E_F)$ is the density of states at the Fermi level per
spin (per unit volume and unit energy). When the quantization axis
is parallel to the local spin polarization, one finds
\begin{equation}
\frac{1}{\rho (E_F)}j_\uparrow   = -D_\uparrow
C-\frac{\widetilde{D}}{l_{sf}}\left[ A\exp(x/l_{sf})-
B\exp(-x/l_{sf})\right] ,
\end{equation}
\begin{equation}
\frac{1}{\rho (E_F)}j_\downarrow = -D_\downarrow
C+\frac{\widetilde{D}}{l_{sf}}\left[ A\exp(x/l_{sf})-
B\exp(-x/l_{sf})\right] ,
\end{equation}
where
\begin{equation}
\widetilde{D}=2\frac{D_\uparrow D_\downarrow}{D_\uparrow +
D_\downarrow}.
\end{equation}

It is convenient to write the spin current in the matrix form as
\begin{equation}
\check{j} = \frac{1}{2}\left[ j_0\check{1}  +
j_z\check{\sigma}_z\right] ,
\end{equation}
with $j_0 = (j_\uparrow +j_\downarrow )$ being the total particle
current density, and $j_z = (j_\uparrow -j_\downarrow )$ being the
total $z$-component of the spin current. Thus, one finds
\begin{equation}
\frac{1}{\rho (E_F)}j_0  = -C(D_\uparrow +D_\downarrow ),
\end{equation}
and
\[
\frac{1}{\rho (E_F)}j_z  =  -C(D_\uparrow -D_\downarrow )
\]
\begin{equation}
-\frac{2\widetilde{D}}{l_{sf}}[A\exp(x/l_{sf})- B\exp(-x/l_{sf})].
\end{equation}
The particle current $j_0$ is related to the charge current $I_0$
{\it via} $I_0=ej_0$, where $e$ is the electron charge ($e<0$).
Thus, positive charge current (flowing from left to right)
corresponds to negative particle current (electrons flow from
right to left).

\subsection{Nonmagnetic films}

Solution of the diffusion equation
 for the distribution functions
inside nonmagnetic films leads to the following equation
\begin{equation}
\check{{\bar\mu}} = \bar{\mu}_0\check{1}  + \mathbf {g}\cdot
\check{\bsig}
\end{equation}
where $\check{\bsig}=(\check{\sigma}_x ,\check{\sigma}_y,
\check{\sigma}_z)$ and in a general case all the three components
of $\mathbf {g}$ are nonzero. The general solutions for
$\bar{\mu}_0$ and $\mathbf {g}$ have the forms
\begin{equation}
\bar{\mu}_0  = Cx+G ,
\end{equation}
\begin{equation}
\mathbf {g} = \mathbf {A} \exp(x/l_{sf})+ \mathbf {B}
\exp(-x/l_{sf}) .
\end{equation}

The spin currents are then given by
\begin{equation}
\check{j} = \frac{1}{2}(j_0\check{1}  + \mathbf{j}\cdot
\check{\bsig}) ,
\end{equation}
with
\begin{equation}
\frac{1}{\rho (E_F)}j_0  = -2CD
\end{equation}
and
\begin{equation}
\frac{1}{\rho (E_F)}\mathbf{j}  =  -\frac{2D}{l_{sf}}[\mathbf{A}
\exp(x/l_{sf})- \mathbf{B}\exp(-x/l_{sf})] ,
\end{equation}
where now $D_\uparrow =D_\downarrow \equiv D$. Of course, all the
constants may be different in different layers.

\subsection{Rotations of the quantization axis}

Distribution function and spin current in the magnetic films are
written in the coordinate system with the axis $z$ along the local
spin polarization. In turn, the formula given above for the
distribution function and spin current inside nonmagnetic films
have general form valid in arbitrary coordinate system. It is
convenient, however, to write them in the system whose axis $z$
coincides with the local quantization axis in one of the adjacent
ferromagnetic films. Since the magnetic moments of the two
ferromagnetic films are non-collinear, it will be necessary to
transform the distribution function and spin current from one
system to another. Thus, if the solution for electrochemical
potentials in a given coordinate system has the form (24), then
the solution in the coordinate system rotated by an angle
$\varphi$ about the axis $x$ is still given by Eq.(24), but with
$\mathbf{g}$ replaced with $\mathbf{g}^\prime$ given by
\begin{equation}
g^{\prime}_x =g_x  ,
\end{equation}
\begin{equation}
g^{\prime}_y =g_y \cos\varphi \; +g_z \sin\varphi\;,
\end{equation}
\begin{equation}
g^{\prime}_z =- g_y \sin\varphi \;+g_z\cos\varphi\; .
\end{equation}
Similar relations also hold when transforming spin current
$\mathbf{j}$ from one coordinate system to the other one.

\section{Boundary conditions and torque}

To determine the unknown constants that enter the general
expressions for electric current and distribution functions inside
all the magnetic and nonmagnetic parts of any layered structure,
we need to specify boundary conditions, which have to be fulfilled
by the distribution function and currents at each interface. Such
boundary conditions were derived by Brataas {\it et al}
\cite{brataas01} within the phenomenological description, and here
we will make use of them.

Charge and spin currents across the normal-metal--ferromagnet
interface (called in the following interfacial currents),
calculated on the normal-metal side in the coordinate system with
the axis $z$ along the local quantization axis in the ferromagnet,
can be written as \cite{brataas01}:
\begin{equation}
e^2j_0 =(G_\uparrow +G_\downarrow )(\bar{\mu}_0^F-\bar{\mu}_0^N)
+(G_\uparrow -G_\downarrow )(g_z^F-g_z^N),
\end{equation}
\begin{equation}
e^2j_z =(G_\uparrow -G_\downarrow )(\bar{\mu}_0^F-\bar{\mu}_0^N)
+(G_\uparrow +G_\downarrow )(g_z^F-g_z^N),
\end{equation}
\begin{equation}
e^2j_x =-2{\rm Re}\{G_{\uparrow\downarrow}\}g_x^N +2{\rm
Im}\{G_{\uparrow\downarrow}\}g_y^N,
\end{equation}
\begin{equation}
e^2j_y =-2{\rm Re}\{G_{\uparrow\downarrow}\}g_y^N -2{\rm
Im}\{G_{\uparrow\downarrow}\}g_x^N,
\end{equation}
where ${\bf g}^N$ (${\bf g}^F$) is the spin accumulation on the N
(F) side of the N/F interface, $G_\uparrow$ and $G_\downarrow$ are
the interfacial conductances in the spin-majority and
spin-minority channels, and $G_{\uparrow\downarrow}$ is the
spin-mixing conductance of the interface, which comes into play
only in non-collinear configurations. It is worth to point that
the above boundary conditions are valid when there is no spin-flip
scattering at the interface.

The boundary conditions can be specified as follows: (i) particle
current is continuous across all interfaces (in all layers and
across all interfaces it is constant and equal to $j_0$), (ii) the
spin current component parallel to the magnetization of a
ferromagnetic layer is continuous across the interface between
magnetic and nonmagnetic layers, and (iii) normal components
(perpendicular to the magnetization of a ferromagnetic film) of
the spin current vanish in the magnetic layer and there is a jump
of these components at the interface between magnetic and
nonmagnetic films, described by Eqs (35) and (36). The above
boundary conditions have to be fulfilled at all interfaces. The
number of the corresponding equations is then equal to the number
of unknown constants, which allows one to determine the spin
accumulation and the charge and spin currents.

Since the perpendicular component of the spin current is absorbed
by the magnetic layers, the corresponding torque $\btau$ per unit
square, exerted on a ferromagnetic film, can be calculated as
\begin{equation}
\btau =\frac{\hbar}{2}\left({\bf j}_{\perp L}-{\bf j}_{\perp
R}\right),
\end{equation}
where ${\bf j}_{\perp L}$ and ${\bf j}_{\perp R}$ are the normal
(to the magnetization) components of the spin current at the left
and right interfaces of the magnetic film, calculated on the
normal metal side of these interfaces. In the simple case (as in
 Fig. 1), where there is no additional magnetic layer outside the
F1/N/F2 trilayer and no transverse spin current at the outer edges
of the trilayer, $\btau$ is simply given by $\btau =-\hbar {\bf
j}_{\perp R}/2$ for F1 and $\btau =\hbar {\bf j}_{\perp L}/2$ for
F2, where ${\bf j}_{\perp R}$ and ${\bf j}_{\perp L}$ have to be
calculated in the nonmagnetic spacer layer (N) at the right
interface of F1 and left interface of F2 (left and right
interfaces of N), respectively.

\section{Torque in a spin-valve structure}

The structure F1/N/F2 under consideration consists of two left
(thick) and right (thin) magnetic films, separated by a
nonmagnetic layer. The thick magnetic film is assumed to be
semi-infinite, so it also plays a role of the left lead. The thin
magnetic film is followed by the right nonmagnetic lead, also
assumed to be semi-infinite. Thickness of the nonmagnetic spacer
layer is $d_0$, whereas of the thin magnetic film is $d_2$. Such a
structure is shown schematically in Fig.1. Both ferromagnetic
films are magnetized in their planes, and magnetization of the
thin layer is rotated by an angle $\varphi$ around the axis $x$
(normal to the films) as shown in Fig.1. Axis $z$ of the
coordinate system is along the net spin of the thick ferromagnetic
film (opposite to the corresponding magnetization). In both
ferromagnetic films the local quantization axes are along the
local net spin, while as the global quantization axis we choose
the local one in the thick ferromagnetic film. According to our
definition, charge current $I_0$ is positive when it flows along
the axis $x$ from left to right, i.e., from the thick towards thin
magnetic films (electrons flow then from right to left).

\begin{figure}
\includegraphics[scale=0.55]{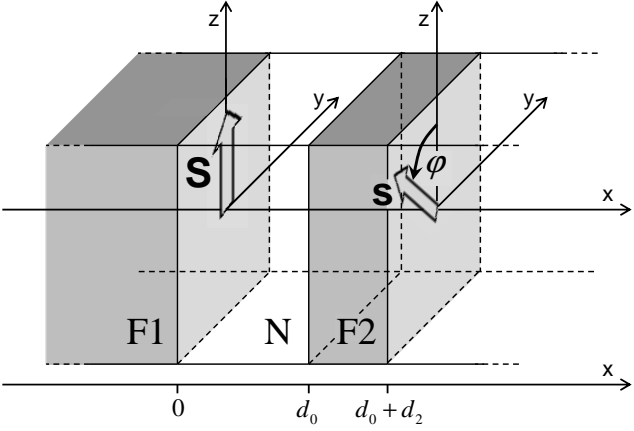}
\caption{Schematic structure of the system studied in this paper.
The system consists of thick (F1) and thin (F2) ferromagnetic
films, separated by a nonmagnetic (N) layer. The thick magnetic
film (similarly as the right nonmagnetic lead) is assumed to be
semi-infinite, while the thin nonmagnetic and ferromagnetic films
have thicknesses $d_0$ and $d_2$, respectively. The arrows
indicate orientation of the net spin of the magnetic films, with
$\varphi$ being the angle between the spins.}
\end{figure}

The in-plane component $\btau_\parallel$ of the torque acting on
the thin magnetic film can be written as
\begin{equation}
\btau_\parallel =a\;\hat{\bf s}\times (\hat{\bf s}\times {\hat
{\bf S}}) ,
\end{equation}
where $\hat{\bf s}$ and $\hat{\bf S}$ are the unit vectors along
the spin polarization of the thin and thick magnetic layers,
respectively. The parameter $a$ is a function of the charge
current $I_0$ (not indicated explicitly in Eq.(38)). Equation (38)
can be rewritten in the form
\begin{equation}
\tau_\varphi =a\,\sin\varphi ,
\end{equation}
where the torque $\tau_\varphi$ is defined in such a way that
positive (negative) torque tends to increase (decrease) the angle
$\varphi$ ($\varphi \in \langle 0 ,2\pi \rangle$) between spin
moments of the films.

The torque $\tau_\varphi$ can be calculated from Eq.(37) as
\begin{equation}
\tau_\varphi = -\frac{\hbar}{2} j_{y}^\prime \vert_{x=d_0}
=-\frac{\hbar}{2} \left( j_{z}\sin \varphi + j_{y}\cos \varphi
\right)\vert_{x=d_0} ,
\end{equation}
where $j_y^\prime$ and $j_{z,y}$ are the components of spin
current in the nonmagnetic thin film written in the local system
of the thin and thick magnetic films, respectively, and calculated
at the very interface between the nonmagnetic an thin magnetic
films. Comparison of Eqs (39) and (40) gives
\begin{equation}
a = -\frac{\hbar}{2\sin\varphi} j_{y}^\prime\vert_{x=d_0}
=-\frac{\hbar}{2} \left( j_{z} + j_{y}\cot \varphi
\right)\vert_{x=d_0}.
\end{equation}

The out-of-plane (normal) component $\btau_{\perp}$ of the torque
may be generally written as
\begin{equation}
\btau_{\perp} =b \; \hat{\bf s}\times \hat{\bf S},
\end{equation}
where the parameter $b$ depends on $I_0$ (not indicated
explicitly). It can be calculated from the formula
\begin{equation}
\tau_x = \frac{\hbar}{2} j_{x}^\prime\vert_{x=d_0}
=\frac{\hbar}{2} j_{x}\vert_{x=d_0},
\end{equation}
where $j_{x}$ ($j_{x}^\prime$) is the $x$-component of the spin
current in the nonmagnetic thin film taken at the interface
between the two thin films and written in the coordinate system of
the thick (thin) magnetic films ($j_x=j_{x}^\prime$ according to
Eq.(30)). From Eqs (42) and (43) follows that the parameter $b$ is
equal
\begin{equation}
b =-\frac{\hbar}{2\sin\varphi} j_{x}^\prime\vert_{x=d_0}
=-\frac{\hbar}{2\sin\varphi} j_{x}\vert_{x=d_0}.
\end{equation}

By taking into account Eqs (35) and (36) one can relate the torque
directly to the spin accumulation in the nonmagnetic film taken at
the interface with the thin magnetic layer. The in-plane (Eq.(40))
and out-of-plane  (Eq.(43)) torque components can be then
rewritten as
\begin{eqnarray}
\tau_\varphi = -\frac{\hbar}{e^2} \left[ {\rm
Re}\{G_{\uparrow\downarrow}\}g_{y}^\prime +{\rm
Im}\{G_{\uparrow\downarrow}\}g_{x}^\prime \right]\vert_{x=d_0}
\nonumber
\\= -\frac{\hbar}{e^2} \left[ {\rm
Re}\{G_{\uparrow\downarrow}\}(g_{y}\cos \varphi +g_{z}\sin
\varphi) \right. \nonumber \\ \left. +{\rm
Im}\{G_{\uparrow\downarrow}\}g_{x} \right]\vert_{x=d_0} ,
\end{eqnarray}
and
\begin{eqnarray}
\tau_x = \frac{\hbar}{e^2} \left[  {\rm
Re}\{G_{\uparrow\downarrow}\}g_{x}^\prime  -{\rm
Im}\{G_{\uparrow\downarrow}\}g_{y}^\prime \right]\vert_{x=d_0}
\nonumber \\
=\frac{\hbar}{e^2} \left[  {\rm Re}\{G_{\uparrow\downarrow}\}g_{x}
-{\rm Im}\{G_{\uparrow\downarrow}\} \right. \nonumber
\\ \left. \times (g_{y}\cos \varphi +g_{z}\sin \varphi
)\right]\vert_{x=d_0}.
\end{eqnarray}

Similarly, the constants $a$ and $b$ can be related to the spin
accumulation {\it via} the formulae
\begin{eqnarray}
a = -\frac{\hbar}{e^2\sin\varphi} \left[ {\rm
Re}\{G_{\uparrow\downarrow}\}g_{y}^\prime +{\rm
Im}\{G_{\uparrow\downarrow}\}g_{x}^\prime \right]\vert_{x=d_0}
\nonumber
\\= -\frac{\hbar}{e^2} \left[ {\rm
Re}\{G_{\uparrow\downarrow}\}(g_{y}\cot \varphi +g_{z}) +{\rm
Im}\{G_{\uparrow\downarrow}\}\frac{g_{x}}{\sin\varphi}
\right]\vert_{x=d_0} ,
\end{eqnarray}
and
\begin{eqnarray}
b = \frac{\hbar}{e^2\sin\varphi} \left[  -{\rm
Re}\{G_{\uparrow\downarrow}\} g_{x}^\prime +{\rm
Im}\{G_{\uparrow\downarrow}\}g_{y}^\prime \right]\vert_{x=d_0}
\nonumber \\
=\frac{\hbar}{e^2} \left[  -{\rm
Re}\{G_{\uparrow\downarrow}\}\frac{g_{x}}{\sin\varphi} +{\rm
Im}\{G_{\uparrow\downarrow}\} (g_{y}\cot \varphi +g_{z}
)\right]\vert_{x=d_0}.
\end{eqnarray}

Equations (40,43) and (41,44) are the final formula for the torque
components and the parameters $a$ and $b$, expressed in terms of
the spin currents. Alternatively, Eqs (45) to (48) are the
corresponding formula expressed in terms of the spin accumulation.
For numerical calculations one can use either the former equations
or equivalently the latter ones.

\section{Numerical results}

For numerical calculations it is convenient to define the bulk and
interfacial spin asymmetry factors for ferromagnetic films
according to the standard definitions \cite{valet93},
\begin{equation}
\rho_{\uparrow (\downarrow )} =2\rho^*(1\mp \beta)
\end{equation}
and
\begin{equation}
R_{\uparrow (\downarrow )} =2R^*(1\mp \gamma) ,
\end{equation}
where $\rho_\uparrow$ and $\rho_\downarrow$ are the bulk
resitivities for spin-majority and spin-minority electrons,
respectively; $R_\uparrow$ and $R_\downarrow$ are the interface
resistances per unit square for spin-majority and spin-minority
electrons, whereas $\beta$ and $\gamma$ are the bulk and
interfacial spin asymmetry coefficients. The formula (49) will
also be used for nonmagnetic films (with $\beta =0$). The
conductances $G_\uparrow$ and $G_\downarrow$ (see Eqs (33,34)) are
then $G_\uparrow=1/R_\uparrow$ and $G_\downarrow=1/R_\downarrow$.
The mixing conductance $G_{\uparrow\downarrow}$ is generally a
complex parameter with the imaginary part being usually one order
of magnitude smaller than the real part.

The key bulk parameters which enter the description, i.e., mean
free paths and diffusion constants can be expressed in a free
electron model by the parameters defined in Eq. (49) and the
relevant Fermi energy $E_F$. In numerical calculations we assume
the same Fermi energy for both magnetic and nonmagnetic layers.
The diffusion parameters $D_{\uparrow (\downarrow )}$ are then
calculated from the formulae (assuming free electron like model
for conduction electrons),
\begin{equation}
D_{\uparrow (\downarrow )}=\frac{1}{3}v_F \lambda_{\uparrow
(\downarrow )},
\end{equation}
where $v_F=\sqrt{2E_F/m_e}$ is the Fermi velocity of electrons,
and the mean free paths $\lambda_{\uparrow (\downarrow )}$ are
\begin{equation}
\lambda_{\uparrow (\downarrow )}=\frac{m_ev_F}{ne^2\rho_{\uparrow
(\downarrow )}},
\end{equation}
with $m_e$ denoting the electron mass and
$n=(1/6\pi^2)(2m_eE_F/\hbar^2)^{3/2}$ being the density of
electrons per spin. Apart from this $\rho (E_F)$ (see Eq.(17)) is
given by $\rho (E_F)=(1/4\pi^2)(2m_e/\hbar^2)^{3/2}E_F^{1/2}$. For
such a description (based on free electron like model) one finds
$\lambda_\downarrow /\lambda_\uparrow =(1-\beta )/(1+\beta )$, and
the parameter $\eta$ defined by Eq.(7) is determined by $\beta$
{\it via} the simple relation
\begin{equation}
\eta =- \beta.
\end{equation}

For nonmagnetic films we use the same definitions, but now
$\lambda_{\uparrow (\downarrow )}$ and $D_{\uparrow (\downarrow
)}$ are independent of the spin orientation (the corresponding
$\beta$ is equal to zero). Finally, the spin diffusion lengths
will be assumed as independent parameters and will be taken from
giant magnetoresistance experiments.

The parameters for the thick ferromagnetic film can be generally
different from those for the thin magnetic film. Similarly,
parameters corresponding to the two nonmagnetic components of the
structure can also be different. In the following, however, we
assume that the nonmagnetic spacer layer (N) and the right
nonmagnetic lead are made of the same material.

The following four different situations have been recently studied
experimentally \cite{darwish04}.

(i)  $\beta_1=\beta_2>0$, $\gamma_1=\gamma_2>0$, which corresponds
to F1 and F2 of the same material with positive spin asymmetries
for both bulk resistivities and interfacial resistances (this
means spin-majority electrons are less scattered both inside the
layers and at the interfaces).

(ii) $\beta_1>0$, $\gamma_1>0$, $\beta_2<0$, $\gamma_2<0$, which
corresponds to different materials for F1 and F2, with positive
spin asymmetries for F1 and negative spin asymmetries for F2.

(iii) $\beta_1=\beta_2<0$, $\gamma_1=\gamma_2<0$, which
corresponds to the same material for F1 and F2, with negative spin
asymmetries for both bulk resisitivities and interfacial
resistances.

(iv) $\beta_1<0$, $\gamma_1<0$, $\beta_2>0$, $\gamma_2>0$, which
corresponds to different materials for F1 and F2, with negative
spin asymmetries for F1 and positive spin asymmetries for F2.

\begin{figure}
\begin{center}
\includegraphics[width=0.7\columnwidth,angle=-90]{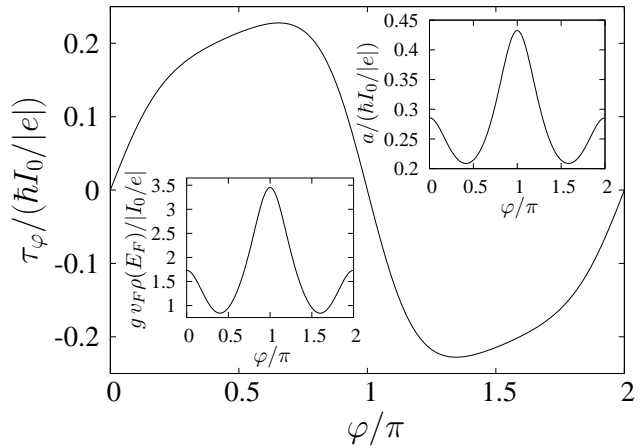}
\caption{Normalized in-plane torque $\tau_\varphi=\vert
\btau_{\parallel}\vert$ acting on the thin ferromagnetic film due
to spin transfer, calculated as a function of the angle $\varphi $
for parameters typical for Co/Cu system, as described in the text.
The insets indicate the similarity between the angular dependence
of the parameter $a$ of the expression $\btau_\parallel
=a\,\hat{\bf s}\times (\hat{\bf s}\times {\hat {\bf S}})$ and that
of the spin accumulation amplitude $g$ (note that $g$ is
normalized to $\vert I_0\vert$ whereas $a$ to $I_0$). The other
parameters are: $E_F=7$eV, $d_0=10$ nm, $d_2=10$ nm.}
\end{center}
\end{figure}

One of the systems within the category (i) is Co/Cu structure,
that has been extensively studied experimentally. For the bulk
resistivities and the interface resistances we take the
experimental values obtained from the GMR measurements
\cite{bass99}. Accordingly, for the Co layers we assume
$\rho^*_1=\rho^*_2=5.1\;\mu\Omega$cm, $\beta_1=\beta_2=0.51$,
$l_{sf}^{(1)}=l_{sf}^{(2)}=60$nm, whereas for the nonmagnetic Cu
layers we assume $\rho_0^*=0.5\;\mu\Omega$cm, $l_{sf}^{(0)}=10^3$
nm.

In turn, for the Co/Cu interfaces we assume $R^*_1=R^*_2=0.52
\cdot 10^{-15}\; \Omega$m$^2$ and $\gamma_1=\gamma_2=0.76$. In
principle, the corresponding mixing conductance
$G_{\uparrow\downarrow}$ could be derived from the angular
dependence of the CPP-GMR. However, in practice there is a large
uncertainty in this derivation and, according to the
experimentalists who have performed this type of experiment
\cite{pratt,dauguet96}, there is no reliable experimental
information on $G_{\uparrow\downarrow}$ from GMR. Therefore, we
assume the value calculated in a free-electron model corrected by
certain factors taken from  {\it ab-initio} calculations by Stiles
\cite{stiles}. For free electron gas and no reflection at the
interface (we assumed the same Fermi energy in all layers), one
finds the following relation between the spin current $j_y^\prime$
and the spin accumulation $g_y^\prime$ components (written in the
coordinate system of the thin magnetic film and taken in the
nonmagnetic spacer at the interface with the sensing (F2) magnetic
film); $j_y^\prime =\rho (E_F)v_Fg_y^\prime /2 =
2G_{\uparrow\downarrow}^{\rm Sh}g_y^\prime /e^2$. Here,
$G_{\uparrow\downarrow} ^{\rm Sh}$ is the Sharvin mixing
conductance in the limit when there is no reflection at the
interface, $G_{\uparrow\downarrow} ^{\rm Sh}=e^2k_F^2/4\pi h$,
with $k_F$ the Fermi wavevector corresponding to the Fermi energy
$E_F$. Reflection from the interface can be taken into account
effectively {\it via} a phenomenological parameter $Q$, writing
${\rm Re}\{G_{\uparrow\downarrow}\}=QG_{\uparrow\downarrow} ^{\rm
Sh}$. In the case of Co/Cu system this factor is roughly equal to
0.925 according to Stiles{\cite{stiles}}. Thus, in the following
numerical calculations we assume ${\rm
Re}\{G_{\uparrow\downarrow}\}=0.542\cdot 10^{-15}\; \Omega$m$^2$.
As for the imaginary part, ${\rm Im}\{G_{\uparrow\downarrow}\}$,
we determine it assuming the same ratio ${\rm
Im}\{G_{\uparrow\downarrow}\}/{\rm Re}\{G_{\uparrow\downarrow}\}$
as that following from {\it ab initio} calculations. Thus, we
assume ${\rm Im}\{G_{\uparrow\downarrow}\}=0.016\cdot 10^{-15}
\;\Omega$m$^2$.

Numerical results on the in-plane and out-of-plane components of
the torque as well as on the corresponding parameters $a$ and $b$
are shown in Fig.2 and Fig.3, respectively. The torque and the
corresponding parameters $a$ and $b$ are normalized to $\hbar I_0
/\vert e\vert$. Within the linear model assumed here the spin
accumulation and spin currents are linear functions of the charge
current, so the curves in Fig.2 are the same for arbitrary
magnitude of the charge current $I_0$. Note that the sign of
torque changes when $I_0$ is reversed.

\begin{figure}
\begin{center}
\includegraphics[width=0.7\columnwidth,angle=-90]{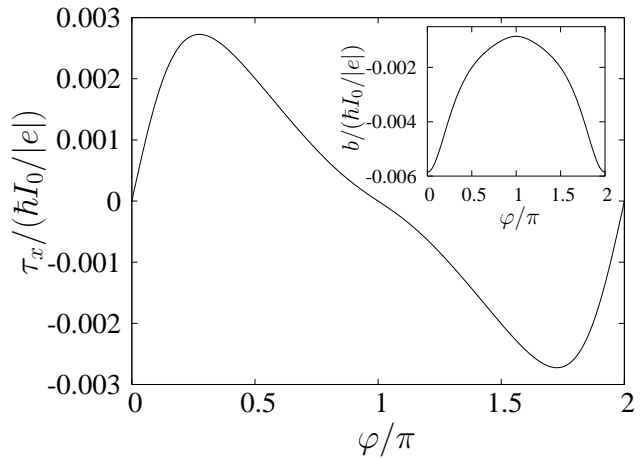}
\caption{Normalized out-of-plane torque, calculated as a function
of the angle $\varphi $ for the parameters typical for Co/Cu
system. The inset shows the corresponding parameter $b$. The other
parameters are the same as in Fig.2.}
\end{center}
\end{figure}

Figure 2 implies that a positive current ($I_0 > 0$) tends to
destabilize the parallel configuration and can switch it to
antiparallel one above a certain threshold value. On the other
hand a negative current tends to destabilize the antiparallel
configuration. This behavior can be defined as a normal
current-induced magnetic switching \cite{darwish04}. The numerical
calculation of the switching currents is not the subject of our
study here. We only note that, by using standard expressions of
the switching currents as a function  of the torque,
magnetization, applied field, anisotropy field, Gilbert
coefficient and thickness of the thin magnetic layer, one  finds
switching currents of the order of $10^7$ A/cm$^2$. What we want
to emphasize is the physical picture emerging from the plots of
Fig.2. The main point is the similar angular dependence of the
spin accumulation amplitude $g=\vert \bf g\vert$ in the
nonmagnetic spacer and of the coefficient $a$ in the expression
(39) for the torque. This results from the relation between the
transverse spin current and spin accumulation in the boundary
conditions involving the mixing conductance, Eq.(36-37). The
interpretation of the angular dependence is straightforward. The
spin accumulation $g$ is larger in the antiparallel configuration,
when the spin direction predominantly transmitted by the thick
layer is slowed down by the thin layer, i.e. $g(\varphi =\pi ) >
g(\varphi =0)$. However, $g(\varphi  )$ does not increase
monotonously from $g(\varphi =0 )$ to $g(\varphi =\pi )$; it
begins with a decrease before getting at $g(\varphi =\pi )$. This
initial decrease follows from the enhanced relaxation of the spin
accumulation due to the efficient pumping of transverse spins as
$\varphi$ departs from zero. This transverse pumping decreases to
zero when $\varphi$ tends to $\pi$ and $g$ goes up to its maximum
value $g(\varphi =\pi )$. The prefactor $a = \tau_\varphi
/\sin\varphi$ follows the same type of variation, with simply a
slightly smaller initial decrease that can be explained by
arguments related to the orientation of $\bf g$, $\hat{\bf s}$ and
$\hat{\bf S}$ (this will appear more clearly in section VI).
Finally $ \tau_\varphi$ varies as $a\sin\varphi$, starting from
zero as $a(\varphi =0)\varphi$ and going back to zero at $\varphi
= \pi$ as $a(\varphi=\pi )(\pi - \varphi )$, that is with a
steeper slope if $a(\varphi =\pi ) > a(\varphi =0)$. The sort of a
shoulder seen in Fig.2 for $\varphi$ slightly below $\pi /2$ is
related to the minimum in $g$ and $a$ at about this angle.

The perpendicular component of the torque, that is the component
coming from the imaginary part of $G_{\uparrow\downarrow}$, is
shown in Fig.3, and is rather small, much smaller than the
in-plane component (it would vanish for ${\rm
Im}\{G_{\uparrow\downarrow}\} =0$). Therefore, it plays a
negligible role in the switching phenomenon, at least for
Co/Cu(111). Although for other interfaces ${\rm
Im}\{G_{\uparrow\downarrow} \}/{\rm Re}\{G_{\uparrow\downarrow}
\}$ is not as small as for Co/Cu(111), the imaginary part of the
mixing conductance leads generally to perpendicular component of
the torque which is definitely smaller than the in-plane one.
Therefore, in the following we will deal only with the in-plane
torque.

Let us consider now the remaining three cases described above;
(ii), (iii) and (iv). For simplicity, we assume the same absolute
values of the bulk and interface spin asymmetry coefficients
$\beta$ and $\gamma$, but their signs are adapted to each
situation. The other parameters are the same for all the cases.
The torque corresponding to the four situations is shown in Fig.4.

The solid curve in Fig.4(a)  corresponds to the case (i), i.e.
$\beta_1 =\beta_2>0$, $\gamma_1= \gamma_2>0$. This curve is the
same as that shown in Fig.2, so that we will not discuss it here
more. Let us change now the sign of the spin asymmetry parameters
of the thin magnetic layer (case (ii) with $\beta_2< 0, \gamma_2 <
0$). The corresponding torque, shown by the dashed line in
Fig.4(a), has the same sign as in the case (i), so that the
switching is still normal ($I_0>0$ destabilizes the parallel
state). With opposite spin asymmetry coefficients in F1 and F2,
the spin accumulation $g$ is larger in the parallel  ($\varphi  =
0$) state than in the antiparallel ($\varphi =\pi$) one, so that
the torque now starts from zero at $\varphi = 0$ with a slope that
is steeper than the slope corresponding to the return point to
zero at $\varphi = \pi$.

The solid line of Fig.4(b) corresponds to the case (iii) with
negative values of all the spin asymmetry coefficients. Compared
to Fig.4(a), the sign of the torque is now reversed, which means
an inversion of the switching currents, as observed in
FeCr/Cr/FeCr structures \cite{darwish04}. As the spin accumulation
$g$ in systems with similar materials for F1 and F2 is larger in
the antiparallel configuration, the slope is higher at the point
where the torque comes back to zero at $\varphi = \pi$.

Finally, for the dashed line of Fig.4(b) corresponding to the case
(iv) ($\beta_1 < 0, \gamma_1<0, \beta_2 > 0, \gamma_2> 0$), the
switching is also inverse. This corresponds to the case studied in
Ref.[\onlinecite{darwish04}] with F1 (F2) and N corresponding to
NiCr (permalloy) and Cu, respectively. Now, the spin accumulation
is larger in the parallel state, so the torque approaches zero at
$\varphi =0$ and $\varphi =\pi$, with the slope larger in the
former case (at $\varphi = 0$).

\begin{figure}
\begin{center}
\includegraphics[width=1.0\columnwidth,angle=-90]{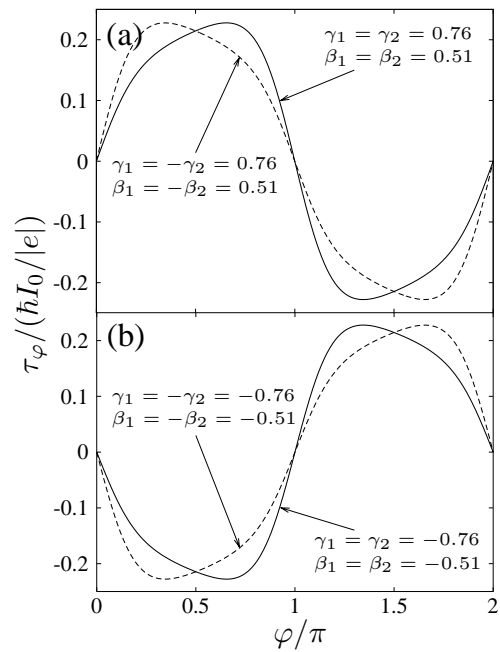}
\caption{Torque as a function of $\varphi $ for the four
situations described in the text, and for indicated values of the
spin asymmetry parameters. The other parameters are the same as in
Fig.2}
\end{center}
\end{figure}

\begin{table*}
\begin{tabular}{|c|c|c|c|c|c|}
\hline
& \multicolumn{2}{|c|}{Spin asymmetries of magnetic layers} &
\multicolumn{2}{|c|}{Sign of torque ($0<\varphi<\pi$)} & \\
\cline{2-5} Situation & \hspace*{1em} Thick layer \hspace*{1em} &
\hspace*{1em} Thin layer \hspace*{1em} & \hspace*{1em} Torque
\hspace*{1em} & \hspace*{1em} Type \hspace*{1em} &
\hspace*{.5em} Sign of GMR \hspace*{.5em} \\
\hline\hline\hline 
(i)   & $(\beta, \gamma) > 0$ & $(\beta, \gamma) > 0$ & $\tau > 0$ & normal & GMR $>$ 0 (normal) \\ 
(ii)  & $(\beta, \gamma) > 0$ & $(\beta, \gamma) < 0$ & $\tau > 0$ & normal & GMR $<$ 0 (inverse) \\
(iii) & $(\beta, \gamma) < 0$ & $(\beta, \gamma) < 0$ & $\tau < 0$ & inverse & GMR $>$ 0 (normal) \\
(iv)  & $(\beta, \gamma) < 0$ & $(\beta, \gamma) > 0$ & $\tau < 0$ & inverse & GMR $<$ 0 (inverse) \\
\hline
\end{tabular}
\caption{Characteristics of the current induced switching in the
four cases studied in this paper. Correlation between the sings of
the torque and GMR is also given.}
\end{table*}

Basic characteristics of the switching phenomena in all the four
cases are gathered in Table I, where the sign of the torque for
positive current ($I_0>0$) is given for $\varphi$ in the range
$0<\varphi <\pi$. The normal/inverse switching phenomenon is
correlated there with the sign of the corresponding GMR effect.

From the results described above follows that it is the spin
asymmetry of the thick magnetic film, which determines whether the
switching effect is normal or inverse. When this spin asymmetry is
positive (negative), one finds a normal (inverse) switching
phenomenon. It is also interesting to note that when the spin
asymmetries of both magnetic films have the same sign, the
structure shows normal GMR effect, whereas when they are opposite,
the corresponding GMR effect is inverse as shown in many CPP-GMR
measurements \cite{vouille99}. Experimental examples of the four
behaviours of Table I can be found in the F1/N/F2 trilayers of
Ref[6] respectively for  NiFe/Cu/NiFe (i), NiFe/Cu/NiCr (ii),
FeCr/Cr/FeCr (iii) and NiCr/Cu/NiFe (iv).

\begin{figure}
\begin{center}
\includegraphics[width=0.8\columnwidth,angle=-90]{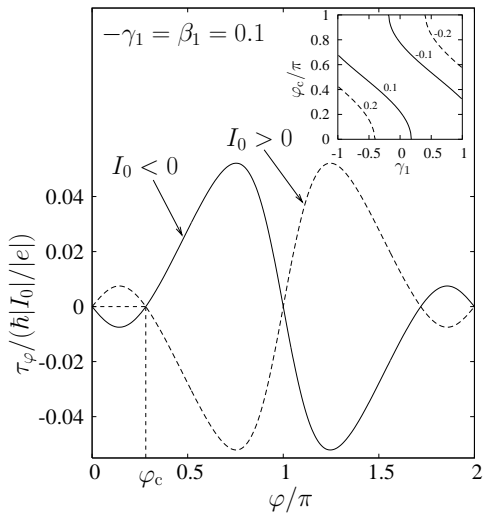}
\caption{Torque for the case when the spin asymmetry factors of
the thick ferromagnetic film, $\beta_1 = 0.1$ and $\gamma_1 =
-0.1$, are significantly different from those of the thin one,
$\beta_2 = 0.51$ and $\gamma_2 = 0.76$. The torque is now
normalized to $\vert I_0 \vert$ and the two curves correspond to
$I_0>0$ and $I_0<0$, as indicated.  The angle $\varphi_c$
corresponds to the point where the torque due to spin transfer
vanishes. Inset shows variation of the angle $\varphi_c$ with the
spin asymmetry factor $\gamma_1$ for several values of $\beta_1$
indicated on the curves. The other parameters are as in Fig.2.}
\end{center}
\end{figure}

For the parameters used in numerical calculations described above
the current-induced torque vanishes in collinear configurations
and one of them (either parallel or antiparallel) is unstable.
This leads to either normal or inverse switching phenomena. An
interesting situation can occur when the amplitudes of spin
asymmetry in the thick and thin layers are different. In Fig. 5,
we show an example of torque calculated for $\beta_1 =0.1$ and
$\gamma_1 =-0.1$, $\beta_2 = 0.51$ and $\gamma_2 = 0.76$. Let us
first consider the case with $I_0 > 0$. The torque is positive
when $\varphi$ increases from zero, then comes back to zero at
$\varphi = \varphi_c$ and becomes negative between  $\varphi_c$
and $\pi$. This means that the torque tends to destabilize both
the parallel and antiparallel states. Above some threshold value
of the current for the instability of the parallel and
antiparallel states, the only solution is a steady state
precession (in the absence of anisotropy and demagnetizing field,
there would be a stable equilibrium at an intermediate orientation
between 0 and $\pi$ in the layer plane, but the general solution
is a precession).

In trilayers which have been studied up to now, steady precessions
with generation of microwave oscillations have been generally
observed when a magnetic field is applied and in a given range of
current density \cite{kiselev03}. The possibility of obtaining
microwave oscillations at zero field would be of great interest
for several devices. Xiao et al \cite{xiao04} have predicted that
microwave oscillations can be observed at zero field and in a
given range of current density when, for asymmetric structures,
the spin transfer and damping torques have markedly different
angular dependences, so that their sum has a wavy angular
dependence. In our structure of Fig.5, the existence of steady
precessions at zero field has a different origin. It comes from
the wavy angular dependence of the torque itself. This structure,
with $I_0 >0$ would be of interest for the generation of
oscillations at zero field and at any value of the current above
some threshold value.

The behavior of Fig.5 for $I_0 <0$ has a different interest. Now,
above some threshold value of the current, the spin transfer
torque stabilizes both the parallel and antiparallel states of the
trilayer. In other words, it increases the damping of the system
in both configurations. This is of interest for some devices, for
example for stabilizing the configuration of read heads based on
CPP-GMR against fluctuations or, at least, avoiding spin transfer
induced fluctuations.

\section{Discussion in the limiting case of real mixing conductance}

As it has been already mentioned before, the imaginary part of the
mixing conductance is usually small. When ${\rm Im}
\{G_{\uparrow\downarrow}\}=0$, the formulae (45) and (46) for the
torque components acquire a simpler form. First of all, the
out-of-plane components of the spin accumulation and spin current
vanish, $g_x=0$ and $j_x=0$. The in-plane torque can be then
written in the form
\begin{eqnarray}
\tau_\varphi = -\frac{\hbar}{e^2}
G_{\uparrow\downarrow}g_{y}^\prime \vert_{x=d_0} \nonumber
\\= -\frac{\hbar}{e^2}G_{\uparrow\downarrow}(g_{y}\cos \varphi +g_{z}\sin \varphi
) \vert_{x=d_0} ,
\end{eqnarray}
where $G_{\uparrow\downarrow}$ is real. In turn, the out-of-plane
torque vanishes then exactly,
\begin{equation}
\tau_x = 0.
\end{equation}

When the spin accumulation in the layer N at its interface with
the layer F2 forms an angle $\varphi_g$ with the axis $z$, then
one finds $g^\prime_y=g\sin (\varphi -\varphi_g )$ and the torque
may be written in the form
\begin{equation}
\tau_\varphi = -\frac{\hbar}{e^2}G_{\uparrow\downarrow}g \sin
(\varphi -\varphi_g ),
\end{equation}
where $g$ is the absolute value (amplitude) of the spin
accumulation at the interface. According to Eq.(41) the parameter
$a$ may be then expressed in the form
\begin{equation}
a = -\frac{\hbar}{e^2}G_{\uparrow\downarrow}g \frac{\sin (\varphi
-\varphi_g )}{\sin\varphi},
\end{equation}

\begin{figure}
\begin{center}
\includegraphics[height=7cm,angle=-90]{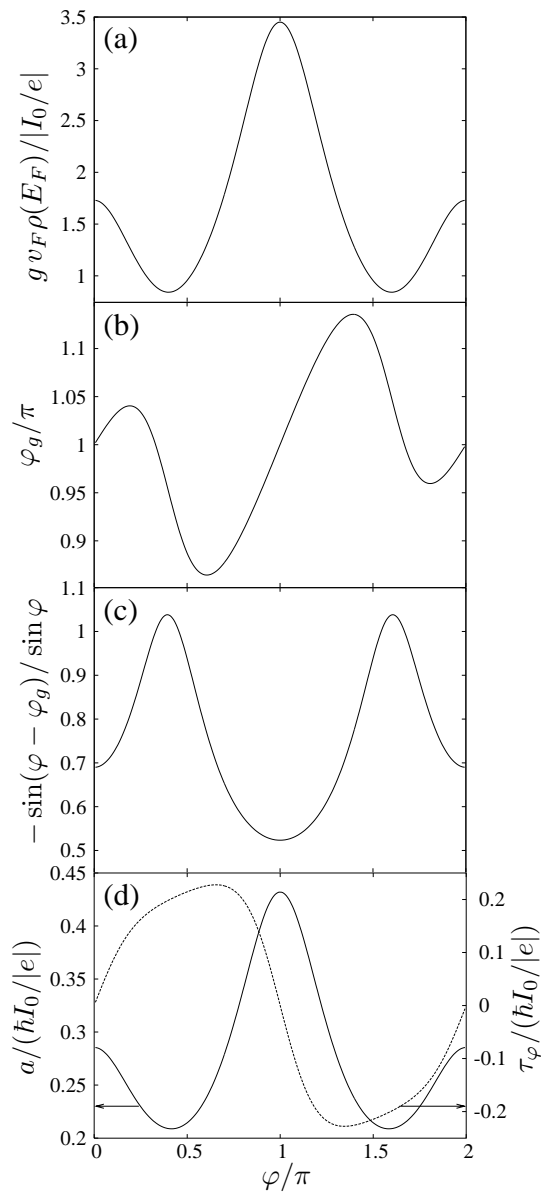}
\caption{Amplitude of spin accumulation at the interface (a), the
angle $\varphi_g$ (solid line in (b)), the factor $\sin (\varphi
-\varphi_g )/\sin\varphi$ (dashed line in (b)), the torque
$\tau_\varphi$ (dashed line in (c)), and the parameter $a$ (solid
line in (c)). All curves are calculated for ${\rm Im}
G_{\uparrow\downarrow}=0$ and for positive current, $I_0>0$. The
other parameters are as in Fig.2. }
\end{center}
\end{figure}

The formulae (56) and (57) relate the torque and the parameter $a$
to the amplitude $g$ of spin accumulation in the layer N at its
interface with F2, and to the {\it sinus} of the angle ($\varphi -
\varphi_g$) between the spin accumulation $\bf g$ and the
polarization axis of the thin ferromagnetic layer F2. It is
interesting to look at the angular variation of these parameters.
We have calculated them for the case (i) and in Fig.6 we show the
angular variation of the spin accumulation amplitude $g$, the
angle $\varphi_g$, the factor $-\sin(\varphi
-\varphi_g)/\sin\varphi$ of Eq.(57), the torque amplitude
$\tau_\varphi$, and the prefactor $a$ of the expression
$\btau_{\parallel}= a\, \hat{\bf s}\times (\hat{\bf s}\times {\hat
{\bf S}})$. All the curves are shown there for positive current,
$I_0>0$. The amplitude $g$ of the spin accumulation, see Fig.6(a),
goes from its small value in the parallel state to its higher
value in the antiparallel state, as expected when F1 and F2 have
the same spin asymmetries. As discussed in section V, the initial
decrease before the upturn to $g$ in the antiparallel state is due
to the relaxation enhancement generated by transverse spin pumping
in a non-collinear state. As for the orientation of $\bf g$, one
can see from the variation of $\varphi_g$ in Fig.6(b) that the
vector $-\bf g$ has first an intermediate orientation between
those of $\hat{\bf S}$ and $\hat{\bf s}$, following the rotation
of $\hat{\bf s}$ at an angle of about $0.7\varphi$ from $\hat{\bf
s}$. Then $-\bf g$ comes back to the orientation of $\hat{\bf S}$
and, as $\varphi$ tends to $\pi$, the angle between $-\bf g$ and
$\hat{\bf S}$ tends to 0 as $-0.55(\pi -\varphi )$. The result of
this angular variation of the spin accumulation is that the factor
$-\sin (\varphi -\varphi_g )/\sin\varphi$ does not change
significantly and varies only between about 0.52 and 1.05 for the
system we have considered, as shown in Fig.6(c). This indicates
that the variation of the torque prefactor $a$, shown in Fig.6(d),
is mainly controlled by the variation of the spin accumulation
amplitude $g$, with a small additional influence of the factor
$-\sin (\varphi -\varphi_g )/\sin\varphi$ (this influence
explains, for example, that the dip around $\varphi = 0.45\pi$ is
less pronounced for $a$ than for $g$). Finally the variation of
$\tau_\varphi $ with the angle $\varphi$, dashed curve in
Fig.6(d), reflects the variation of the product of $a$ times
$\sin\varphi$.

We have seen in section V that similar angular variation of $g$
and $a =\tau_\varphi /\sin\varphi$ also occurs in the general case
with nonzero but small imaginary part of $G_{\uparrow\downarrow}$.
The explanation is the same as above, but only a little more
complex because $\bf g$ has also a normal component. The general
conclusion is that the torque is closely related to the spin
accumulation in the nonmagnetic layer N. We put forward two
practical consequences of this correlation. (1) The ratio between
the switching current amplitudes of the
parallel$\rightarrow$antiparallel and
antiparallel$\rightarrow$parallel transitions, which reflects
approximately the ratio between the derivatives $d\tau_\varphi
/d\varphi$ at $\varphi = 0$ and $\varphi=\pi$, simply reflects the
ratio between $g$ in the parallel and antiparallel configurations.
In particular this ratio is inverted when the sign of the spin
asymmetries is inverted in one of the magnetic layers, as in the
situations (iii) and (iv) in section V. (2) More generally, the
torque amplitude can be enhanced or reduced by enhancing or
reducing the spin accumulation in the nonmagnetic spacer. This has
been confirmed by the experiments  of Ref.[\onlinecite{darwish04}]
in which the torque could be enhanced or reduced by introducing
spin-flip scatterers at different places in the structure.

\section{Comparison with the previous calculation of the torque for $\varphi$ close to $0$ or $\pi$ }

A simple expression of the current-induced torque has been derived
by Fert et al \cite{fert04} in the small angle limit, that is when
the angle $\varphi$ between the two magnetic layers ($\theta$ in
the notation of Ref.[\onlinecite{fert04}]) is small or close to
$\pi$. This expression involves also parameters derived from
CPP-GMR data and has been recently fitted with experimental
results of CIMS on samples doped in several ways \cite{darwish04}.
It is interesting to compare the torques at small angle calculated
with this expression and by using the model of the present
article. For simplicity, we will compare the two calculations when
the mixing conductance is real, which allows us a clearer physical
picture of the differences between the two approaches. With zero
imaginary mixing conductance ($\epsilon = 0$ in the notation of
Ref.[\onlinecite{fert04}]), the expression of the torque at small
angle is given by Eq.(5) of Ref.[\onlinecite{fert04}],
\[
\tau =\hbar\left[\left(\frac{v_Fm_N^{\rm
P(AP)}}{8}+\frac{j_{m,N}^{\rm
P(AP)}}{2}\right)(1-e^{-d_0/\lambda}) \right.
\]
\begin{equation}
\left.+\left(\frac{v_Fm_{F1}^{\rm P(AP)}}{4}+j_{m,F1}^{\rm
P(AP)}\right)e^{-d_0/\lambda}\right] \hat{\bf s}\times (\hat{\bf
s}\times {\hat {\bf S}}) ,
\end{equation}
where $m_N^{\rm P(AP)}(j_{m,N}^{\rm P(AP)})$  is the spin
accumulation density (spin current ) in the N layer at the N/F2
interface in the parallel (P) and antiparallel (AP)
configurations, $m_{F1}^{\rm P(AP)}(j_{m,F1}^{\rm P(AP)})$ are the
same quantities in F1 at the F1/N interface ($m$ and $j_m$ are
defined as positive for polarizations in the majority spin
direction of F1, note also that, in the notation of
Ref.[\onlinecite{fert04}], one electron counts for ±1/2 in $m$ and
$j_m$) and $\lambda$ is the mean free path in N. The torque we
have calculated in the preceding sections corresponds to the first
and dominant term of this expression in the limit $d_0 >> \lambda$
(we will come back later to the meaning of the other terms).
Omitting the last three terms, Eq.(58) can be written as
\begin{equation}
\tau =\hbar\frac{v_Fm_N^{\rm P(AP)}}{8} \hat{\bf s}\times
(\hat{\bf s}\times {\hat {\bf S}}) .
\end{equation}

The spin accumulation density $m$ can be written as a function of
the spin accumulation g of our paper in the following way:
\begin{equation}
m=\frac{m_ek_F}{2\pi^2\hbar^2}g
\end{equation}
so that Eq.(59) can be written as
\begin{equation}
\tau =\frac{\hbar}{2e^2}G^{Sh}_{\uparrow\downarrow}g \; \hat{\bf
s}\times (\hat{\bf s}\times {\hat {\bf S}}) ,
\end{equation}
to be compared with the torque found in section VI:
\begin{equation}
\tau =-\frac{\hbar}{e^2}G_{\uparrow\downarrow} g\frac{\sin(\varphi
-\varphi_g)}{\sin\varphi}\hat{\bf s}\times (\hat{\bf s}\times
{\hat {\bf S}}) .
\end{equation}
The only differences between Eqs.(61) and (62) are:

(i) The replacement of $G^{Sh}_{\uparrow\downarrow}$ by
$G_{\uparrow\downarrow}$. In section V, we assumed
$G_{\uparrow\downarrow}$ equal to 0.925$\times
G^{Sh}_{\uparrow\downarrow}$ for Co/Cu interface. The factor 0.925
is also the factor $t$ which was approximated by 1 between Eq.(4)
and Eq.(5) in Ref.[\onlinecite{fert04}]).

(ii) The replacement of the factor $1/2$ by the value of
$-\sin(\varphi -\varphi_g)/\sin(\varphi)$ for $\varphi$ close to 0
and $\pi$ - approximately by 0.7 and 0.52 in the case illustrated
in Fig.6(b). Actually the factor $1/2$ in Eq.(61) comes from the
values $\varphi_g=\pi + \varphi /2$ for $\varphi$ infinitesimally
small ($\theta_m =\theta /2$ in the notation of
Ref.[\onlinecite{fert04}]) and $\varphi_g =\pi-(\pi-\varphi )/2$
for $\pi-\varphi$ infinitesimally small. These values of
$\varphi_g$ are derived from a transverse spin conservation
condition with the assumption of a constant orientation of $\bf g$
in a thin enough nonmagnetic layer. Our calculations in this paper
show that, for a Cu layer of 10 nm, this assumption does not
strictly hold and that the factor 0.5 must be replaced by 0.7 and
0.52 for $\varphi$ close respectively to 0 and $\pi$.

We thus conclude that the torques expressed respectively by
Eq.(61), that is derived from the first term in the expression of
the torque in Ref.[\onlinecite{fert04}], and by Eq.(62) derived in
this article, differ only by a numerical factor not very different
from unity. This factor will tend to unity for thinner N and will
depart more from unity for thicker layer N.

It remains to discuss why our calculations do not include the
three last terms of Eq.(58). This comes from the mixing
conductance approximation \cite{brataas00}, that is the
approximation of the boundary equations for the transverse
components, Eqs.(35,36). These equations express the diffusion
transverse spin current generated by the discontinuity between the
finite value of ${\bf g}_\perp$ (${\bf g}_\perp$ denotes the spin
accumulation component normal to the magnetization) in N and its
zero value in F. The finite value of ${\bf g}_\perp$ is taken just
at the interface, which assumes that the gradient of ${\bf
g}_\perp$ and the resulting variation of  ${\bf g}_\perp$ on a
distance of the order of the mean free path can be neglected. The
second term of Eq.(58) takes into account the contribution from
this gradient to the diffusion current. In addition, if N is
thinner than the mean free path, a certain amount of the diffusion
current comes directly from F1, which gives rise to the last two
terms of Eq.(58). In conclusion, the calculation of the torque at
small angle in this article has the advantage of a more accurate
determination of the orientation of the spin accumulation in the
nonmagnetic spacer, that is, for example, a more accurate
determination of the factor $\sin(\varphi -\varphi_g)/\sin\varphi$
involved in Eq.(62). In the system we considered, the difference
is relatively small (for a 10 nm thick spacer, 5\% and 40\%  for
$\varphi$ close respectively to 0 and $\pi$), and it should
decrease (increase) for thinner (thicker layers). On the other
hand, with the boundary conditions of Eqs.(35,36), we are not able
to calculate the contributions to the diffusion current due to the
gradient of spin accumulation and to the direct diffusion from the
thick magnetic layers. These contributions, corresponding to the
last three terms of Eq.(58), must be taken into account for thin
nonmagnetic spacers -- the diffusion from the thick layer vanishes
only when the spacer is thicker than the mean free path. We will
introduce them in a further extension of our model.

\section{Concluding remarks}

In conclusion, we have presented a model of CIMS which is partly
based on the classical transport equations derived from the
Boltzmann equation standard model of CPP-GMR \cite{valet93}.
Additional boundary equations based on the concept of mixing
conductance are used to describe the interfacial absorption of
transverse spin currents in non-collinear magnetic configurations.
This model applied to Co/Cu/Co trilayers  allows us to calculate
the spin transfer torques as a function  of the usual parameters
derived from CPP-GMR measurements (interface spin asymmetry
coefficients and resistances, bulk spin  asymmetry coefficients
and resistivities, spin diffusion lengths), and the mixing
conductance coefficients derived from ab initio calculations.  We
have also shown that the torque and its angular dependence is
closely related to the spin accumulation in the nonmagnetic spacer
and to its angular dependence. Enhancing the spin accumulations
seems to be the way to reduce the critical currents. The model has
been also applied to situations with different spin asymmetries in
the two magnetic layers of F1/N/F2 structures to reproduce the
inversion of the switching current recently obtained by reversing
the spin asymmetry of the thick magnetic layer. By applying it to
asymmetric structures, we have shown that steady precessions can
be obtained in zero field. We have also pointed out some
application limits for boundary conditions and indicated that
certain corrections can be anticipated by going beyond these
limits.

\begin{acknowledgments}
This work is partly supported by Polish State Committee for
Scientific Research through the Grants No.~PBZ/KBN/044/P03/2001,
No.~2~P03B~053~25, and by the "Spintronics" RT Network of the EC
RTN2-2001-00440 and FCT Grant No. POCTI/FIS/58746/2004. One of us
(AF) thanks Vincent Cros, Julie Grollier and Henri Jaffres for
very fruitful discussions. We also thank Mark Stiles for having
communicated to us numerical data concerning the transmission of
transverse spin currents at the Co/Cu(111) interface.
\end{acknowledgments}


\begin{thebibliography}{}

\bibitem{slonczewski96}
J. C. Slonczewski, J. Magn. Magn. Mater. {\bf 159}, L1 (1996);
{\bf 195}, L261 (1999).

\bibitem{berger96} L. Berger, Phys. Rev B {\bf 54}, 9353 (1996).

\bibitem{tsoi98} M. Tsoi, A.G.M. Jansen, J. Bass, W.-C. Chiang, M.
Seck, V. Tsoi, and P. Wyder, Nature (London) {\bf 406}, 46 (1998);
E.B. Myers, D.C Ralph, J.A. Katine, R.N. Louie, and R.A. Buhrman,
Science {\bf 285}, 867 (1999); J.E. Wegrowe, D. Kelly, T. Truong,
Ph. Guittienne, and J.Ph. Ansermet, Europhys. Lett {\bf 56}, 748
(2001).

\bibitem{sun99} J. A. Katine, F.J. Albert, R.A. Buhrman,
E.B. Myers, and D.C. Ralph, Phys. Rev. Lett. {\bf 84}, 3149
(2000); J. Grollier, V. Cros, A. Hamzic, J.M. George, H. Jaffres,
A. Fert, G. Faini, J. Ben Youssef, and H. Legall, Appl. Phys.
Lett. {\bf 78}, 3663 (2001); J.Z. Sun, D.J. Monsma, D.W. Abraham,
M.J. Rooks, and R.H. Koch, Appl. Phys. Lett. {\bf 81}, 2202
(2002).

\bibitem{urazhdin04} S. Urazhdin, N. O. Birge, W. P. Pratt, and J. Bass, Appl. Phys.
Lett. 84, 1516, (2004).

\bibitem{darwish04}
M.AlHajDarwish, H. Kurt, S. Urazhdin, A. Fert, R. Loloee, W.P.
Pratt Jr., and J. Bass, Phys. Rev. Lett., {\bf 93}, 157203 (2004).

\bibitem{baibich88}  M.N. Baibich, J.M. Broto, A. Fert, F. Nguyen van Dau, F.
Petroff, P. Etienne, G. Creuzet, A. Friederich and J. Chazelas,
Phys. Rev. Letters {\bf 61}, 2472 (1988); G. Binasch, P.
Gr\"unberg, F. Saurenbach, and W. Zinn, Phys. Rev B {\bf 39}, 4828
(1989); J. Barna\'s, A. Fuss, R.E. Camley, P. Gr\"unberg and W.
Zinn, Phys. Rev. B {\bf 42}, 8110 (1990).

\bibitem{moodera95} J. S. Moodera, L. R. Kinder, T. M. Wong and R. Meservey, Phys.
Rev. Lett. 74, 3273 (1995).

\bibitem{kiselev03} S. I. Kiselev, J. C. Sankey, I. N. Krivorotov, N. C. Emley,
R. J. Schoelkopf, R. A. Buhrman and D. C. Ralph, Nature 425, 380 (2003); S. I.
Kiselev, J. C. Sankey, I. N. Krivorotov, N. C. Emley, M. Rinkoski, C. Perez,
R. A. Buhrman, and D. C. Ralph, Phys. Rev. Lett. 93, 036601 (2004) ; W. H.
Rippard, M. R. Pufall, S. Kaka, S. E. Russek, and T. J. Silva, Phys. Rev.
Lett. 92, 027201 (2004); W. H. Rippard, M. R. Pufall, S. Kaka, T. J. Silva,
and S. E. Russek, Phys. Rev. B 70, 100406(R) (2004).

\bibitem{weintal00}
X. Waintal E.B. Myers, P.W. Brouwer, and D.C. Ralph, Phys. Rev. B
{\bf 62}, 12317 (2000); Yu.A Bazaliy, B.A. Jones, and S.C. Zhang,
Phys. Rev. B {\bf 57}, R3213 (1998); J.E. Wegrowe, Phys. Rev. B
{\bf 62}, 1 (2000); P. Weinberger, A. Vernes, B.L. Györphy and L.
Szunyogh, Phys. Rev. B 70, 094401 (2004); D.M. Edwards, F.
Frederici, J. Mathon and A. Umerski, cond-mat/0407562.

\bibitem{slonczewski02} J. Slonczewski, J. Magn. Magn. Mat.. 247, 324 (2002); L. Berger,
J. Appl. Phys. 89, 5521 (2001); S. Zhang, P. M. Levy, and A. Fert,
Phys. Rev. Lett. {\bf 88}, 236601 (2002); A. Shpiro, P. M. Levy,
and S. Zhang, Phys. Rev. B {\bf 67}, 104430 (2003).

\bibitem{stiles02}
M.D. Stiles and A. Zangwill, Phys. Rev. B {\bf 66}, 014407 (2002);
J. Appl. Phys. 91, 6812 (2002).

\bibitem{brataas00}
A. Brataas, Yu.V. Nazarov, and G.E.W. Bauer, Phys. Rev. Lett. {\bf
84}, 2481 (2000); D.H. Hernando, Yu.V. Nazarov, A. Brataas, and
G.E.W. Bauer, Phys. Rev. B {\bf 62}, 5700 (2000); Y. Tserkovnyak
and A. Brataas Phys. Rev. B  {\bf 65}, 094517 (2002); Y.
Tserkovnyak,  A. Brataas, and G.E.W. Bauer, Phys. Rev. Lett., {\bf
88}, 117601 (2002); Y. Tserkovnyak,  A. Brataas, G.E.W. Bauer, and
B.I. Halperin, cond-mat/0409242; A. A. Kovalev, A. Brataas, and G.
E. W. Bauer, Phys. Rev. B {\bf 66}, 224424 (2002).

\bibitem{valet93}
T. Valet, A. Fert, Phys. Rev. B, {\bf 48}, 7099 (1993).

\bibitem{bass99}
J. Bass and W. P. Pratt Jr., J. Magn. Magn. Mat. 200,  274 (1999).

\bibitem{brataas01}
A. Brataas, Yu. V. Nazarov, G. E. W. Bauer, Eur. Phys. J. B {\bf
22}, 99, (2001).

\bibitem{xia01}
K. Xia, P. J. Kelly, G. E. W. Bauer, A. Brataas, and I. Turek,
Phys. Rev. B, {\bf 65}, 220401 (2002); M. Zwierzycki, Y.
Tserkovnyak, P. J. Kelly, A. Brataas, and G. E. W. Bauer,
arXiv:cond-mat/0402

\bibitem{fert04}
A. Fert, V. Cros, J.M. George, J. Grollier, H. Jaffr\`{e}s, A.
Hamzic, A. Vaur\`{e}s, G. Faini, J. Ben Youssef, and H. Le Gall,
J. Magn. Magn. Metals (2004).

\bibitem{pratt} W.P. Pratt, private communication

\bibitem{dauguet96} P. Dauguet, P. Gandit, J. Chaussy, S.F. Lee,
A. Fert, P. Holody, Phys. Rev. B54, 1083 (1996).

\bibitem{stiles}
M.D. Stiles, private communication.

\bibitem{vouille99} C. Vouille, A. Barth\`{e}l\`{e}my, A. Fert, P.A. Schroeder, S. Hsu,
A. Reilly, R. Loloee, Phys. Rev. B 60, 6710 (1999).

\bibitem{xiao04} J. Xiao, A. Zangwill, and M.D. Stiles, Phys. Rev. B 70, 172405 (2004).


\end{thebibliography}
\end{document}